\documentclass[conference]{IEEEtran}
\IEEEoverridecommandlockouts
\usepackage{cite}
\usepackage{amsmath,amssymb,amsfonts}
\usepackage{algorithmic}
\usepackage{graphicx}
\usepackage{textcomp}
\usepackage{xcolor}
\usepackage{multirow}

\def\BibTeX{{\rm B\kern-.05em{\sc i\kern-.025em b}\kern-.08em
    T\kern-.1667em\lower.7ex\hbox{E}\kern-.125emX}}
\begin{document}

\title{On the Incentive Compatibility of\\ Block Propagation in Bitcoin}

\author{
\IEEEauthorblockN{Fumichika Maeda\textsuperscript{*}}
\IEEEauthorblockA{
\textit{Kyoto University}\\
Kyoto, Japan \\
fumichika.research@gmail.com
}
\and
\IEEEauthorblockN{Akira Sakurai\textsuperscript{*}}
\IEEEauthorblockA{
\textit{Kyoto University}\\
Kyoto, Japan \\
akilucky0304@gmail.com
}
\and
\IEEEauthorblockN{Taishi Nakai}
\IEEEauthorblockA{
\textit{Kyoto University}\\
Kyoto, Japan \\
distributed.nakai@gmail.com
}
\and
\IEEEauthorblockN{Kazuyuki Shudo}
\IEEEauthorblockA{
\textit{Kyoto University}\\
Kyoto, Japan \\
shudo@computer.org
}
\thanks{\textsuperscript{*}The first two authors contributed equally to this work.}
}

\maketitle

\begin{abstract}
Bitcoin is permissionless and does not rely on any central administrator, which gives it strong censorship resistance.
At the same time, it is important to incentivize miners to behave in ways that align with the interests of the system as a whole.
This paper asks whether miners are individually incentivized to propagate blocks, one of the most fundamental processes in Bitcoin.
Miners collectively maintain the blockchain by generating blocks and disseminating them across the network.
If miners have an incentive not to propagate some blocks, this would indicate a fundamental flaw in Bitcoin’s incentive design. 
Although prior work has studied how propagation delays affect forks and mining rewards, it has not fully characterized miners’ incentives to improve block propagation under different tie-breaking rules.
To address this gap, we derive analytical reward expressions for each tie-breaking rule based on a blockchain network model that captures the effect of forks on mining fairness.
These expressions explicitly characterize how block propagation delays, hashrate distribution, and tie-breaking rules jointly determine mining rewards.
We then use them to analyze miners’ incentives to improve block propagation.
Our results show, for example, that miners have no mining-reward incentive to relay blocks generated by other miners.
By contrast, under the first-seen rule, every non-majority miner is incentivized to receive other miners’ blocks more quickly and to propagate its own blocks more quickly.
Finally, we compare tie-breaking rules and identify a trade-off between propagation incentives and mining fairness.
In particular, the first-seen rule provides the strongest incentives to reduce propagation delays, but it also worsens mining fairness the most.
\end{abstract}

\begin{IEEEkeywords}
Bitcoin, incentive design, block propagation, mining fairness, tie-breaking rules
\end{IEEEkeywords}

\section{Introduction}

Bitcoin~\cite{bitcoin} is a peer-to-peer monetary system that enables transaction processing without relying on a trusted third party.
Its central feature is that anyone can participate in the system, and no specific administrator controls which participants may join or which transactions should be accepted.
This permissionlessness and the absence of a trusted authority are the basis of Bitcoin's censorship resistance.

At the same time, these properties make incentive design essential.
Because Bitcoin lacks a central administrator and participation is open, the system cannot assume that participants share common objectives or act altruistically for the benefit of the network.
Instead, the behavior of participants must be regulated through incentives.
In other words, the protocol must be designed so that actions desirable for the
system are also rational for each individual participant.

In this paper, we focus on the incentive compatibility of block propagation.
Block propagation is one of the most fundamental processes in Bitcoin.
When a miner generates a new block, the block must be disseminated through the network so that other miners can build on it.
Efficient propagation benefits the system by reducing unintended forks, thereby contributing to security, consistency, and mining fairness~\cite{onTheSecurity, EverythingisaRace, TightConsistencyBoundsforBitcoin, informationpropagation, OnScalingDecentralizedBlockchains}. 
Conversely, propagation delays increase the fork rate and can therefore undermine these properties.

Existing studies on block-propagation incentives rely on simplified models and therefore provide only limited insights~\cite{LessIsMore, WhenisSlower}.
To overcome this problem, we build on the blockchain network model proposed by Sakurai et al.~\cite{sakurai2025modelbasedanalysisminingfairness}.
This model captures the effect of block propagation delay on mining rewards under different tie-breaking rules.
However, the model is difficult to use directly for incentive analysis.
In particular, the relationships among hashrate distribution, propagation delay, and mining rewards are not analytically transparent.
As a result, it is hard to identify from the model alone when a miner benefits from improving or worsening a specific propagation delay.

To address this limitation, we derive theoretical reward expressions from the
model.
These expressions explicitly describe how each miner's expected reward depends
on hashrate distribution, block propagation delays, and tie-breaking rules.
We then use the resulting formulas to analyze block-propagation incentives.
Our analysis characterizes the conditions under which such incentives arise.
Finally, we compare tie-breaking rules from the perspective of block-propagation incentives.

\paragraph*{Contributions}

\begin{itemize}
    \item We derive analytical reward expressions that explicitly capture how block propagation delays, hashrate distribution, and tie-breaking rules affect mining rewards. We validate these expressions through simulation-based experiments. The resulting expressions provide a unified analytical framework for studying how network-level factors affect mining rewards.

    \item We analyze block-propagation incentives by classifying delays into relay, outbound, and inbound delays. We show that miners have no mining-reward incentive to reduce relay delays for blocks generated by other miners. We also derive necessary and sufficient conditions under which miners are incentivized to reduce outbound and inbound delays under each tie-breaking rule. Finally, we formulate feasible delay-manipulation strategies in Bitcoin and analyze their system-level effects.

    \item We compare representative tie-breaking rules from the perspective of block-propagation incentives. The first-seen rule provides the strongest incentives to reduce propagation delays among the analyzed rules. However, stronger propagation incentives can conflict with mining fairness. This reveals a trade-off between propagation incentives and reward fairness in the design of tie-breaking rules.
\end{itemize}

\section{Bitcoin}\label{sec:bitcoin} In Bitcoin~\cite{bitcoin}, users initiate transactions by broadcasting them over the peer-to-peer network. These transactions are collected into blocks by nodes called miners.

Each block references a single parent block, and this referencing structure forms a chain of blocks, called the blockchain. Each miner maintains its own local view of the block tree and selects, as its main chain, the chain with the greatest cumulative proof of work. This rule is often referred to as the longest-chain rule. 

To generate a block, a miner constructs a block template containing information such as the processed transactions, a timestamp, and the hash of the latest block on its main chain. The miner then repeatedly computes hashes by changing the nonce and other modifiable fields until the hash of the block header falls below a specified target. This process is called mining. A block satisfying this condition is considered valid and is propagated through the network. After verifying the validity of the block, each miner updates its own local view of the blockchain. Only blocks included in the main chain contribute to the accepted transaction history, and the miner who generated such a block receives a block reward. This reward consists of a block subsidy and transaction fees. 

An adversary can invalidate a processed transaction by constructing a competing chain that excludes the block containing the target transaction. Specifically, the adversary attempts to create a fork and make its competing chain overtake the chain that includes the target transaction. The success probability of such an attack depends on the adversary's hashrate share and the number of confirmations. As long as the adversary controls less than a majority of the total hashrate, the probability of catching up decreases as confirmations accumulate. In this probabilistic sense, Bitcoin makes transaction reversal increasingly difficult. 

Even when all miners follow the Bitcoin protocol honestly, forks can still occur during block propagation. A fork occurs when a new block is generated before a previously generated block has fully propagated through the network. The likelihood of such a fork increases with block propagation delay~\cite{informationpropagation}. In this paper, we focus on unintentional forks caused by propagation delays and do not consider intentional forks caused by adversarial mining strategies. 

When multiple competing chains have the same cumulative proof of work, miners need a tie-breaking rule to decide which chain to mine on. In Bitcoin, miners typically mine on the chain containing the competing block they received first. We refer to this policy as the first-seen rule. This rule is simple and requires no additional coordination among miners. However, from the perspective of selfish mining, it has a drawback: the security of the system can become sensitive to the adversary's block propagation capability~\cite{majorityisnotenough}. To address this issue, alternative tie-breaking rules have been proposed, such as the random rule~\cite{majorityisnotenough}, which selects a competing chain at random, and the last-generated rule~\cite{sakurai2024fullylocallastgeneratedrule, sakurai2024tiebreaking, oneweirdtricktostopselfishminers}, which selects the chain containing the most recently generated block.
\section{Related Work}\label{related work} 

The line of work most closely related to our study examines the relationship between propagation delays, forks, and mining rewards.
Decker et al. showed that block propagation delays affect mining rewards through unintentional blockchain forks~\cite{informationpropagation}.
Mao et al. showed that, in some settings, reducing a miner's connectivity can increase its block reward~\cite{LessIsMore}.
However, their model abstracts away several features of realistic blockchain networks. For example, their analysis treats the block generation interval as a deterministic fixed value.
Lu et al. analyzed conditions under which larger delays can increase a miner's  block reward~\cite{WhenisSlower}. However, their results are limited in scope, because their model considers only a uniform outbound delay for the miner's own blocks.
Our work extends this line of block-propagation incentive analysis in several directions.
First, we use a more detailed blockchain network model, distinguish among relay, inbound, and outbound delays, and analyze their incentives.
Second, we compare how different tie-breaking rules affect block propagation incentives.

Our work primarily studies incentive compatibility in block propagation.
At the same time, incentive compatibility in Bitcoin has also been studied from
other perspectives, especially in the context of strategic mining.
Prior work has shown that deviations from honest mining, such as selfish mining and block withholding, can be profitable~\cite{majorityisnotenough,rosenfeld2011analysis,subversiveminerstrategiesblock,FAW,OptimalSelfishMining,onTheSecurity,OntheInstability}.
By contrast, Kiayias et al. showed that honest mining is rational under certain decentralization conditions~\cite{BlockchainMiningGames}.
Unlike these studies, which mainly analyze miners' incentives in block publication or withholding strategies, we focus on whether miners have incentives to improve or manipulate the propagation process itself.

\section{Model}\label{sec:model}

We describe the Bitcoin network model used in this paper.
We first consider the baseline setting in which all miners honestly follow the
Bitcoin protocol.
Let $V$ denote the set of miners.
Each miner $i \in V$ has a hashrate share $\alpha_i$, satisfying
$\sum_{i \in V} \alpha_i = 1$.
Let $T$ denote the expected block generation interval.
We assume that these values remain fixed throughout the analysis.
We also assume that the block reward is constant across all blocks.

We analyze mining fairness using the mining profit rate
($\mathrm{MPR}$).
The mining profit rate of miner $i$ is defined as
\begin{align}
    \mathrm{MPR}(i)
    =
    \frac{r_i - \alpha_i}{\alpha_i},
    \label{eq:mpr-definition}
\end{align}
where $r_i$ denotes the block reward proportion of miner $i$.

To capture the effect of forks on the mining profit rate, we use the notion of
a round, following Sakurai et al.~\cite{sakurai2025modelbasedanalysisminingfairness}.
A round is a global time interval defined as the period from the generation
of the first block at height $h$ to the generation of the first block at
height $h+1$.
We assume that at most two blocks are generated in a single round.
This round-based model enables us to handle forks formally and capture their
impact on the mining profit rate.

A fork is defined as the event in which two blocks are generated within a
single round.
Let $F_{ij}$ denote the probability that a fork occurs when miner $i$ initiates
a round and miner $j$ is the next block generator.
Then $F_{ij}$ is given by $F_{ij}=1 - \exp(-T_{ij}/T)$,
where $T_{ij}$ is the time from the generation of a block by miner $i$ to its
reception and validation by miner $j$.

Let $W_{ij}$ denote the probability that, in such a fork, the block generated
by miner $i$ is included in the main chain.
The value of $W_{ij}$ depends on the adopted tie-breaking rule and is given by
\begin{align}
    W_{ij}
    =
    \begin{cases}
        \displaystyle \sum_{k \in V} \alpha_k p_{i,j,k}
        & \text{(first-seen rule)},\\
        \displaystyle \alpha_i + \frac{1-\alpha_i-\alpha_j}{2}
        & \text{(random rule)},\\
        \displaystyle \alpha_i
        & \text{(last-generated rule)}.
    \end{cases}
    \label{eq:wij}
\end{align}
Here, $p_{i,j,k}$ denotes the probability that miner $k$ receives miner $i$'s block before miner $j$'s block when miner $i$ starts the round and miner $j$ causes a fork.
It is given by
\begin{align}
    p_{i,j,k}
    =
    \begin{cases}
        1
        & \text{if } T_{ik} < T_{jk},\\
        0
        & \text{if } T_{ik} > T_{ij} + T_{jk},\\
        \displaystyle
        \frac{
        \exp\left(-\frac{T_{ik}-T_{jk}}{T}\right)
        -
        \exp\left(-\frac{T_{ij}}{T}\right)
        }{
        1 - \exp\left(-\frac{T_{ij}}{T}\right)
        }
        & \text{otherwise}.
    \end{cases}
    \label{eq:pijk}
\end{align}

\section{Analytical Expressions for Mining Fairness}
\label{sec:analytical-framework}

In this section, we derive the relationship between $\mathrm{MPR}$ and network
parameters such as block propagation delays and hashrate distributions based on
the model described in Section~\ref{sec:model}.
We first present the assumptions used in the analysis.
We then define aggregate delay quantities that capture how quickly each miner
can propagate and receive blocks, and we summarize the analytical results.
Detailed derivations are provided in the appendix.
Finally, we validate the analytical expressions through simulation experiments.

\subsection{Assumptions}
\label{sec:assumptions}

We make two additional assumptions.
First, block propagation delays are sufficiently small compared with the block
generation interval.
Second, block propagation delays satisfy a propagation-based triangle
inequality.

\paragraph*{Cond.~A: First-order approximation in the small-delay regime}

We assume that the block propagation delay between any pair of miners is
sufficiently small compared with the block generation interval.
Specifically,
\begin{align}
\varepsilon_{ij}
:=
\frac{T_{ij}}{T}
\ll 1
\quad
\text{for all } i,j \in V .
\label{eq:cond-a}
\end{align}
We then apply a first-order approximation with respect to the small parameters
$\varepsilon_{ij}$.
That is, we neglect all second- and higher-order terms, including products such
as $\varepsilon_{ij}\varepsilon_{kl}$.

\paragraph*{Cond.~B: A propagation-based triangle inequality}

For any miners $i,j,k \in V$, we assume that
\begin{align}
T_{ik} \leq T_{ij} + T_{jk}.
\label{eq:cond-b}
\end{align}
This assumption is used only in the theoretical analysis of the first-seen rule.

This condition can be interpreted as follows.
Suppose miner $i$ generates a block and miner $j$ obtains it before miner $k$
does, i.e., $T_{ij} \leq T_{ik}$.
Then $T_{ik}-T_{ij}$ represents the remaining time until miner $k$ finishes
validating the same block after miner $j$ has already done so.
Cond.~B states that this remaining time is at most $T_{jk}$, the time required
for a block generated by miner $j$ to reach miner $k$.
This is natural because, once miner $j$ has validated miner $i$'s block, the
block may already be known by multiple miners and can continue propagating
toward miner $k$ from several sources.
By contrast, when miner $j$ generates a new block, only miner $j$ knows it
initially.
Therefore, the former time is typically no larger than the latter.

An exception may arise under adversarial network conditions, especially
eclipse-style attacks.
In a classical eclipse attack, an adversary monopolizes a victim's incoming and
outgoing connections and can thereby filter or delay the victim's view of the
blockchain~\cite{eclipseattack}.
More recent work has also shown that block propagation to a victim can be
significantly delayed through eclipse-based manipulation of the victim's
neighbors~\cite{TendrilStaller}.
In such situations, miner $k$ may first receive a block advertisement through
adversarially controlled neighbors, while the corresponding block delivery is
intentionally delayed.
Then, even after miner $j$ has already validated miner $i$'s block, miner $k$
may still experience a remaining delay larger than $T_{jk}$, and hence Cond.~B
can fail.
For this reason, Cond.~B should be understood as a benign-network assumption
used for tractable analysis, rather than as a property expected to hold under
active network-layer attacks.

\subsection{Analytical Expressions}
\label{sec:analytical-expressions}

We now present the analytical expressions used for the mining-fairness
analysis.
First, we define the following aggregate inbound and outbound delay quantities:
\begin{align}
f_{\mathrm{in}}^{(m)}(i)
&:=
\sum_{j \in V} (\alpha_j)^m T_{ji},
\label{eq:aggregate-inbound-delay}
\\
f_{\mathrm{out}}^{(m)}(i)
&:=
\sum_{j \in V} (\alpha_j)^m T_{ij},
\label{eq:aggregate-outbound-delay}
\\
\bar{f}
&:=
\sum_{i \in V} \alpha_i f_{\mathrm{in}}^{(1)}(i)
=
\sum_{i \in V} \alpha_i f_{\mathrm{out}}^{(1)}(i)
=
\sum_{i \in V,\, j \in V} \alpha_i \alpha_j T_{ij}.
\label{eq:average-delay}
\end{align}

Under Cond.~A and Cond.~B, the analytical expressions for $\mathrm{MPR}(i)$
under the first-seen, random, and last-generated rules are as follows:
\begin{align}
\mathrm{MPR}(i)
&=
\frac{1}{T}
\bigl(
2\bar{f}
-
f_{\mathrm{in}}^{(1)}(i)
-
f_{\mathrm{out}}^{(1)}(i)
\bigr),
\tag{FS}
\label{eq:mpr-first-seen}
\\
\mathrm{MPR}(i)
&=
\frac{1}{T}
\bigl\{
\bar{f}
-
\frac{1-\alpha_i}{2}
\bigl(
f_{\mathrm{in}}^{(1)}(i)
+
f_{\mathrm{out}}^{(1)}(i)
\bigr)
\nonumber\\
&\quad
-
\frac{1}{2}
\bigl(
f_{\mathrm{in}}^{(2)}(i)
+
f_{\mathrm{out}}^{(2)}(i)
\bigr)
\bigr\},
\tag{RD}
\label{eq:mpr-random}
\\
\mathrm{MPR}(i)
&=
\frac{1}{T}
\bigl\{
\bar{f}
-
(1-\alpha_i)f_{\mathrm{out}}^{(1)}(i)
-
f_{\mathrm{in}}^{(2)}(i)
\bigr\}.
\tag{LG}
\label{eq:mpr-last-generated}
\end{align}
Here, \eqref{eq:mpr-first-seen}, \eqref{eq:mpr-random}, and
\eqref{eq:mpr-last-generated} correspond to the first-seen, random, and
last-generated rules, respectively.

Detailed derivations are provided in the appendix for each tie-breaking rule.

\subsection{Simulation Validation}
\label{sec:simulation-validation}

The analytical expressions above hold under Cond.~A and Cond.~B.
In this subsection, we assess their practical applicability by comparing them
with model-based calculations using blockchain-network parameters obtained from
simulations.

\subsubsection{Validation Method}
\label{sec:validation-method}

Our validation compares two quantities:
$\mathrm{MPR}^{\mathrm{ref}}$, which is computed directly from the underlying
model, and $\mathrm{MPR}^{\mathrm{approx}}$, which is obtained from the
analytical expressions derived in this paper.
To construct realistic validation settings, we used the blockchain-network
simulator \textit{SimBlock}~\cite{simblock} to obtain block propagation delays
in simulated networks together with the corresponding hashrate distributions of
miners.
Using these values, we computed both $\mathrm{MPR}^{\mathrm{ref}}$ and
$\mathrm{MPR}^{\mathrm{approx}}$ and compared them.

As evaluation metrics, we used Lin's concordance correlation coefficient
(CCC)~\cite{CCC} and the relative error.
CCC measures not only correlation but also agreement between the two quantities.
The relative error measures the magnitude of the approximation error relative
to $\mathrm{MPR}^{\mathrm{ref}}$ and is defined as
\begin{align}
\delta_{\mathrm{approx}}
&=
\frac{
\left\|
\mathrm{MPR}^{\mathrm{approx}}
-
\mathrm{MPR}^{\mathrm{ref}}
\right\|_2
}{
\left\|
\mathrm{MPR}^{\mathrm{ref}}
\right\|_2
}.
\label{eq:approx-relative-error}
\end{align}

We do not directly estimate $\mathrm{MPR}$ from \textit{SimBlock} by counting
mining rewards over long simulation runs, because such a validation becomes
computationally impractical at the network scale considered in this paper.
Sakurai et al.~\cite{sakurai2025modelbasedanalysisminingfairness} validated
their model using a relatively small network of only 10 nodes, precisely because
larger networks make such direct validation unrealistic.
They reported that, for each validation target, 50 simulation runs were
performed in parallel, and that each combination of block propagation delay and
block generation interval required approximately four days of computation,
whereas the corresponding mining-fairness computation based on the model was
completed within several tens of
milliseconds~\cite{sakurai2025modelbasedanalysisminingfairness}.
This computational gap makes direct simulation-based estimation of
$\mathrm{MPR}$ unsuitable for large-scale validation.

This limitation does not undermine the purpose of our validation.
Our goal is not to estimate realized mining rewards directly from long-run
simulations, but to evaluate how accurately the analytical expressions
reproduce the model-based $\mathrm{MPR}$ values under realistic network
parameters.

\subsubsection{Simulation Setting}
\label{sec:simulation-setting}

We used \textit{SimBlock}~\cite{simblock} to generate the propagation-delay
matrix $[T_{ij}]$ used in the validation.
For each setting, we first constructed a network topology, assigned node roles
and hashrates, and then measured the time required for a block generated by
miner $i$ to reach miner $j$.
The resulting matrix $[T_{ij}]$ was then used to compute both
$\mathrm{MPR}^{\mathrm{ref}}$ and $\mathrm{MPR}^{\mathrm{approx}}$.

The simulated network was divided into six geographic regions.
Each node was assigned to one region, and the delay of each communication link
was determined by the regional latency, the transmission time implied by the
link bandwidth, and the receiver-side verification delay.
We used the 2024 latency and bandwidth parameters implemented in
\textit{SimBlock}.
As the block-propagation protocol, we employed compact block relay
(CBR)~\cite{bip152}.
Since \textit{SimBlock} natively implements only the low-bandwidth mode of CBR,
we additionally implemented the high-bandwidth mode.
For CBR nodes, the probability of successful block reconstruction was controlled
by the reconstruction-success parameter provided by
\textit{SimBlock}~\cite{IdentifyingImpactsofProtocolandInternetDevelopmentontheBitcoinNetwork}.

We considered the following two settings.

\paragraph*{Realistic setting}

The first setting was designed to approximate a Bitcoin-like network.
The network consisted of 24{,}000 nodes~\cite{bitnodes}: 20 pool miners,
980 solo miners, 11{,}000 reachable non-miners, and 12{,}000 unreachable
non-miners.
Only pool miners and solo miners performed mining; all non-miners were assigned
zero hashrate.

The hashrate distribution of the 20 pool miners was set according to an
empirical Bitcoin-like distribution~\cite{MiningPoolStats}.
The remaining mining power was assigned to the 980 solo miners using an
empirical solo-mining distribution constructed from public pool statistics, so
that every solo miner had positive hashrate while remaining substantially
smaller than major pools.

The public-layer degree distribution was specified primarily according to the
empirical measurements of Grundmann et al.~\cite{PeerDegreeDistribution}.
Each pool miner was assigned total degree 125.
Based on prior measurement results, we did not give pool miners any additional
advantage in terms of connection degree~\cite{miller2015topology}.
For solo miners and reachable non-miners, total degrees were sampled from an
empirically inspired reachable-node distribution that had a strong mass near
125, a broad lower tail, and a thin high-degree tail.
Each unreachable non-miner was assigned degree 10, corresponding to an
outbound-only client.
Subject to these degree constraints, public-layer neighbors were selected
uniformly at random.

To model the practical relay infrastructure used primarily by large pools, we
added a relay network on top of the public P2P layer.
The relay-network implementation was based on the design proposed by
Otsuki et al.~\cite{relayNetwork}.
One relay server was placed in each geographic region, and each pool miner was
connected to the relay server in its own region.
The bandwidth of both pool-miner--relay links and inter-relay links was set to
ten times that of an ordinary public P2P link.
When a pool miner generated a block, the block was first sent to its regional
relay server, then disseminated among the relay servers, and finally forwarded
from each relay server to the pool miners in the corresponding region.

\paragraph*{Uniform-random setting}

The second setting served as a homogeneous baseline.
It consisted of 1{,}000 miners with identical hashrate.
No relay network was provided in this setting.
Each miner was assigned a total degree of 10, and neighbors were selected
uniformly at random.
Compared with the realistic setting, this baseline removed both hashrate
concentration and pool-specific relay advantage.

\begin{table*}[t]
\centering
\caption{Validation results for block generation intervals of $600$~s and
$60$~s. The table reports the CCC between $\mathrm{MPR}^{\mathrm{ref}}$ and
$\mathrm{MPR}^{\mathrm{approx}}$, the relative error
$\delta_{\mathrm{approx}}
=
\lVert \mathrm{MPR}^{\mathrm{approx}}
-
\mathrm{MPR}^{\mathrm{ref}} \rVert_2
/
\lVert \mathrm{MPR}^{\mathrm{ref}} \rVert_2$,
and the corresponding fork rate.}
\label{tab:validation-results}
\footnotesize
\setlength{\tabcolsep}{3.5pt}
\begin{tabular}{@{}lllrrrl@{}}
\hline
Block interval & Setting & Rule & CCC & $\delta_{\mathrm{approx}}$ & Fork rate & Ref. fork rate \\
\hline
\multirow{6}{*}{$600$\,s}
& \multirow{3}{*}{Realistic}
& First-seen      & 0.998034 & 0.0352576  & \multirow{3}{*}{0.000375431} & \multirow{6}{*}{\shortstack{0.00015\\(Bitcoin)}} \\
& & Random         & 0.999997 & 0.00133235 &  &  \\
& & Last-generated & 0.999996 & 0.00212041 &  &  \\
\cline{2-6}
& \multirow{3}{*}{Uniform}
& First-seen      & 0.999803 & 0.0200159 & \multirow{3}{*}{0.000737312} &  \\
& & Random         & 0.999999 & 0.00167713 &  &  \\
& & Last-generated & 0.999997 & 0.00239832 &  &  \\
\hline
\multirow{6}{*}{$60$\,s}
& \multirow{3}{*}{Realistic}
& First-seen      & 0.997432 & 0.0404288 & \multirow{3}{*}{0.00375334} & \multirow{6}{*}{\shortstack{0.00127\\(Dogecoin)}} \\
& & Random         & 0.999665 & 0.0133518  &  &  \\
& & Last-generated & 0.999619 & 0.0213573  &  &  \\
\cline{2-6}
& \multirow{3}{*}{Uniform}
& First-seen      & 0.999626 & 0.0277126  & \multirow{3}{*}{0.00736887} &  \\
& & Random         & 0.999861 & 0.0168351  &  &  \\
& & Last-generated & 0.999715 & 0.0241587  &  &  \\
\hline
\end{tabular}
\end{table*}

\subsubsection{Validation Results}
\label{sec:validation-results}

Table~\ref{tab:validation-results} shows the validation results for block
generation intervals of $600$~s and $60$~s.
For each setting, the table reports the CCC and the relative error between
$\mathrm{MPR}^{\mathrm{ref}}$ and $\mathrm{MPR}^{\mathrm{approx}}$ under the
three tie-breaking rules, together with the corresponding fork rate.

Overall, $\mathrm{MPR}^{\mathrm{approx}}$ agreed closely with
$\mathrm{MPR}^{\mathrm{ref}}$ in all cases.
When the block generation interval was $600$~s, the CCC was at least
$0.998034$ for every tie-breaking rule and every network setting, indicating
almost perfect agreement.
When the interval was shortened to $60$~s, the fork rate increased by
approximately one order of magnitude.
Nevertheless, the CCC remained at least $0.997432$ even in this shorter-block
regime.
Thus, the analytical approximation remained close to the model-based calculation
even under substantially higher fork rates.

The relative error also remained moderate in all cases.
For the $600$~s interval, the maximum relative error was $0.0352576$.
For the $60$~s interval, the maximum relative error was $0.0404288$.
Although the errors increased when the block interval was shortened, the largest
relative error over all experiments was still only about $4.1$\%.

To quantify the overall error scale, we combine our approximation error with
the simulation--model error reported in previous work on model-based
$\mathrm{MPR}$ estimation~\cite{sakurai2025modelbasedanalysisminingfairness}.
The previous work reported relative errors between
$\mathrm{MPR}^{\mathrm{sim}}$ and $\mathrm{MPR}^{\mathrm{ref}}$ for each
tie-breaking rule.
In the setting corresponding to a fork rate of approximately $0.01$, the
reported errors were $0.0108707$ for the first-seen rule, $0.0164265$ for the
random rule, and $0.0210495$ for the last-generated rule.

Combining these errors with our relative errors between
$\mathrm{MPR}^{\mathrm{ref}}$ and $\mathrm{MPR}^{\mathrm{approx}}$ gives
rule-specific upper bounds on the relative error between
$\mathrm{MPR}^{\mathrm{sim}}$ and $\mathrm{MPR}^{\mathrm{approx}}$.
For the $60$~s interval, the resulting bounds are $0.051739$, $0.029998$, and
$0.042856$ in the realistic setting for the first-seen, random, and
last-generated rules, respectively.
In the uniform setting, the corresponding bounds are $0.038885$, $0.033538$,
and $0.045717$.
Thus, in the fork-rate range closest to the previous reference setting, the
overall relative error is bounded by at most $0.051739$, or about $5.2$\%.

As a more conservative worst-case estimate, we also combine the largest
previously reported simulation--model error for each tie-breaking rule with the
largest approximation error observed for the same rule in our experiments.
This gives worst-case bounds of $0.094610$ for the first-seen rule, $0.120369$
for the random rule, and $0.187714$ for the last-generated rule.
Therefore, even under this conservative rule-wise worst-case combination, the
overall relative error between $\mathrm{MPR}^{\mathrm{sim}}$ and
$\mathrm{MPR}^{\mathrm{approx}}$ is bounded by approximately $18.8$\%.
Because these worst-case bounds combine errors from different experimental
settings, they should be interpreted as conservative estimates rather than
observed errors.

These results show that the analytical expressions derived in this paper can
accurately estimate $\mathrm{MPR}$ both in homogeneous random networks and in
more realistic blockchain network settings.
Furthermore, the fork rates observed in our experiments were higher than the
reference fork rates reported for real-world systems~\cite{Calibratingtheperformance}.
Specifically, for the $600$~s block interval, the observed fork rates were
$0.000375431$ and $0.000737312$, both exceeding the Bitcoin reference fork rate
of $0.00015$.
For the $60$~s block interval, the observed fork rates were $0.00375334$ and
$0.00736887$, both exceeding the Dogecoin reference fork rate of $0.00127$.
This suggests that the analytical approximation remains applicable even in
regimes with higher fork rates than those observed in representative real-world
systems, including systems with short block generation intervals such as
Dogecoin.
Thus, the validation supports the use of the analytical expressions in a broad
range of practical blockchain-network settings.
\section{Incentives for Block Propagation}
\label{sec:propagation-incentives}

We now analyze block propagation from an incentive-compatibility perspective.
As discussed in the introduction, fast block propagation is desirable at the
system level because it reduces unintended forks and contributes to security,
consistency, and mining fairness.
However, this does not imply that each miner has an individual incentive to
reduce propagation delays.
The purpose of this section is to characterize when reducing a given propagation
delay is rational for an individual miner.

Our analysis is based on the total differentials of $\mathrm{MPR}(i)$.
Specifically, by taking the total differentials of the reward expressions in
\eqref{eq:mpr-first-seen}--\eqref{eq:mpr-last-generated}, we obtain the
following expressions.
Let $V_{-i}:=V\setminus\{i\}$.
\begin{align}
d\mathrm{MPR}(i)
&=
\frac{1}{T}
\biggl\{
\sum_{j,k\in V_{-i}}
2\alpha_j\alpha_k\,dT_{jk}
+
\sum_{j\in V_{-i}}
\alpha_j(2\alpha_i-1)\,dT_{ji}
\nonumber\\
&\quad
+
\sum_{k\in V_{-i}}
\alpha_k(2\alpha_i-1)\,dT_{ik}
\biggr\},
\tag{dFS}
\label{eq:dmpr-first-seen}
\\
d\mathrm{MPR}(i)
&=
\frac{1}{T}
\biggl\{
\sum_{j,k\in V_{-i}}
\alpha_j\alpha_k\,dT_{jk}
\nonumber\\
&\quad
+
\frac{1}{2}
\sum_{j\in V_{-i}}
\alpha_j(3\alpha_i-1-\alpha_j)\,dT_{ji}
\nonumber\\
&\quad
+
\frac{1}{2}
\sum_{k\in V_{-i}}
\alpha_k(3\alpha_i-1-\alpha_k)\,dT_{ik}
\biggr\},
\tag{dRD}
\label{eq:dmpr-random}
\\
d\mathrm{MPR}(i)
&=
\frac{1}{T}
\biggl\{
\sum_{j,k\in V_{-i}}
\alpha_j\alpha_k\,dT_{jk}
+
\sum_{j\in V_{-i}}
\alpha_j(\alpha_i-\alpha_j)\,dT_{ji}
\nonumber\\
&\quad
+
\sum_{k\in V_{-i}}
\alpha_k(2\alpha_i-1)\,dT_{ik}
\biggr\}.
\tag{dLG}
\label{eq:dmpr-last-generated}
\end{align}
Here, \eqref{eq:dmpr-first-seen}, \eqref{eq:dmpr-random}, and
\eqref{eq:dmpr-last-generated} correspond to the first-seen, random, and
last-generated rules, respectively.

For miner $i$, we classify propagation delays into three types.
First, a \emph{relay delay} is a delay $T_{jk}$ for a block generated by
another miner $j$ to reach another miner $k$, where $j \neq i$ and $k \neq i$.
This delay captures the propagation of blocks generated by other miners, which
miner $i$ may help reduce by relaying those blocks.
Second, an \emph{inbound delay} is a delay $T_{ji}$ for a block generated by
another miner $j$ to reach miner $i$.
This delay determines how quickly miner $i$ receives blocks generated by other
miners.
Third, an \emph{outbound delay} is a delay $T_{ik}$ for a block generated by
miner $i$ to reach another miner $k$.
This delay determines how quickly blocks generated by miner $i$ are delivered
to other miners.

\paragraph*{Relay delay}

We first consider miner $i$'s incentive to reduce relay delays for blocks
generated by other miners.
The coefficients of $dT_{jk}$ in
\eqref{eq:dmpr-first-seen}--\eqref{eq:dmpr-last-generated}, where $j \neq i$
and $k \neq i$, are
\begin{align*}
    &2\alpha_j\alpha_k
    && \text{under the first-seen rule}, \\
    &\alpha_j\alpha_k
    && \text{under the random rule}, \\
    &\alpha_j\alpha_k
    && \text{under the last-generated rule}.
\end{align*}
Since $\alpha_j>0$ and $\alpha_k>0$ for active miners, all these coefficients
are positive.
Thus, the marginal gain from reducing $T_{jk}$ is negative under all three
tie-breaking rules.
Therefore, miner $i$ has no mining-reward incentive to reduce such relay
delays; indeed, increasing them increases $\mathrm{MPR}(i)$.

This gives a basic negative result for the incentive compatibility of block
propagation.
Although relaying blocks generated by other miners is beneficial for the system
because it improves propagation, it is not individually rational from the
viewpoint of mining rewards.

\paragraph*{Inbound delay}

Next, we consider miner $i$'s incentive to reduce inbound delays for blocks
generated by other miners.
The coefficients of $dT_{ji}$ in
\eqref{eq:dmpr-first-seen}--\eqref{eq:dmpr-last-generated}, for $j \neq i$, are
$\alpha_j(2\alpha_i-1)$ under the first-seen rule,
$\tfrac{1}{2}\alpha_j(3\alpha_i-1-\alpha_j)$ under the random rule,
and $\alpha_j(\alpha_i-\alpha_j)$ under the last-generated rule.
Hence, reducing $T_{ji}$ increases $\mathrm{MPR}(i)$ if and only if
\begin{align*}
    \alpha_i &< \frac{1}{2}
    && \text{under the first-seen rule}, \\
    \alpha_i &< \frac{1+\alpha_j}{3}
    && \text{under the random rule}, \\
    \alpha_i &< \alpha_j
    && \text{under the last-generated rule}.
\end{align*}

These conditions show that inbound block-propagation incentives depend on both
the hashrate distribution and the tie-breaking rule.
Under the first-seen rule, every miner with less than half of the total
hashrate has an incentive to reduce inbound delays for blocks generated by
other miners.
Under the random rule, this incentive is weaker and depends on the hashrate
share $\alpha_j$ of the miner that generated the block.
In particular, a miner with $\alpha_i \leq 1/3$ has an incentive to reduce
$T_{ji}$ for every other active miner $j$, whereas a larger miner may lack this
incentive for some $j$.
Under the last-generated rule, miner $i$ has an incentive to reduce $T_{ji}$
only when the block is generated by a miner with a larger hashrate share than
miner $i$.

\paragraph*{Outbound delay}

Finally, we consider miner $i$'s incentive to reduce outbound delays for its
own blocks.
The coefficient of $dT_{ik}$ in
\eqref{eq:dmpr-first-seen}--\eqref{eq:dmpr-last-generated} is
$\alpha_k(2\alpha_i-1)$ under both the first-seen and last-generated rules,
and $\tfrac{1}{2}\alpha_k(3\alpha_i-1-\alpha_k)$ under the random rule.
Therefore, reducing $T_{ik}$ increases $\mathrm{MPR}(i)$ if and only if
\begin{align*}
    \alpha_i &< \frac{1}{2}
    && \text{under the first-seen rule}, \\
    \alpha_i &< \frac{1+\alpha_k}{3}
    && \text{under the random rule}, \\
    \alpha_i &< \frac{1}{2}
    && \text{under the last-generated rule}.
\end{align*}

Thus, outbound block-propagation incentives also depend on the hashrate
distribution and the tie-breaking rule.
Under the first-seen and last-generated rules, every miner with less than half
of the total hashrate has an incentive to reduce outbound delays for its own
blocks.
Under the random rule, this incentive depends on the hashrate share $\alpha_k$
of the miner to which the block is delivered.
In particular, a miner with $\alpha_i \leq 1/3$ has an incentive to reduce
$T_{ik}$ for every other active miner $k$, whereas a larger miner may lack this
incentive for some $k$.

\paragraph*{Implications}

The above analysis yields three implications.
First, none of the three tie-breaking rules gives miners a mining-reward
incentive to reduce relay delays for blocks generated by other miners.
This result shows a basic incentive-compatibility problem in block propagation:
although reducing such delays is beneficial for the system as a whole, it is not
individually rational for miners from the viewpoint of mining rewards.

Second, the incentives to reduce inbound and outbound delays are not uniform.
They depend on hashrate shares and on the adopted tie-breaking rule.
In general, larger miners have systematically weaker block-propagation
incentives.
In particular, a miner with at least half of the total hashrate has no strict
mining-reward incentive to reduce either inbound or outbound delays under any
of the three tie-breaking rules.

Third, the first-seen rule provides the strongest block-propagation incentives
among the three rules.
For both inbound and outbound delays, every non-majority miner has an incentive
to reduce delays under the first-seen rule.
The random rule weakens both inbound and outbound incentives, while the
last-generated rule preserves the same outbound condition as the first-seen rule
but weakens the inbound condition.

\paragraph*{Incentive incompatibility in real-world mining}

The above analysis is stated in terms of the mining profit rate, but the
incentive-incompatibility result carries over to real-world mining
profitability.
Recall that $\mathrm{MPR}(i)=(r_i-\alpha_i)/\alpha_i$, where $r_i$ is the block
reward proportion of miner $i$.
Equivalently, $r_i=\alpha_i(1+\mathrm{MPR}(i))$.
Therefore, when the hashrate share $\alpha_i$ is fixed, the sign of the
marginal effect of a propagation delay on $r_i$ is the same as the sign of its
marginal effect on $\mathrm{MPR}(i)$.

Let $\Pi_i$ denote the real-world profit of miner $i$.
For a propagation delay $T_{ab}$, where $T_{ab}$ denotes a relay, inbound, or
outbound delay for miner $i$, we decompose $\Pi_i$ locally as
\begin{align}
\Pi_i
&=
N^{\mathrm{blk}} R_i r_i
-
C_i^{\mathrm{base}}
-
C_i^{\mathrm{delay}}(T_{ab}),
\label{eq:real-world-profit}
\end{align}
where $N^{\mathrm{blk}}>0$ is the total number of rewarded blocks generated in
the network during the period under consideration, and $R_i>0$ is the monetary
reward obtained by miner $i$ per rewarded block.
Thus, $N^{\mathrm{blk}}R_i r_i$ represents miner $i$'s monetary mining revenue
over this period.
In the local marginal analysis below, we treat $N^{\mathrm{blk}}$ and $R_i$ as
fixed with respect to the marginal change in $T_{ab}$.

The term $C_i^{\mathrm{base}}$ denotes cost components that are independent of
the marginal change in $T_{ab}$.
It includes factors such as ASIC efficiency, electricity price, hardware cost,
pool fees, and the baseline scale of mining operation when $\alpha_i$ is fixed.
These factors affect the level of real-world profit, but they vanish when we
take the partial derivative with respect to $T_{ab}$.
By contrast, $C_i^{\mathrm{delay}}(T_{ab})$ includes factors that may depend on
propagation-delay improvement, such as bandwidth cost, relay-node operation,
validation capacity, peering arrangements, and other network investments.

We assume that
$\partial C_i^{\mathrm{delay}}/\partial T_{ab} \leq 0$.
This assumption means that reducing a propagation delay is not costless:
achieving a smaller $T_{ab}$ requires additional network, validation, or
operational resources.
Equivalently, not improving the delay weakly reduces these non-reward costs.
Using $r_i=\alpha_i(1+\mathrm{MPR}(i))$, we obtain
\begin{align}
\frac{\partial \Pi_i}{\partial T_{ab}}
=
N^{\mathrm{blk}} R_i \alpha_i
\frac{\partial \mathrm{MPR}(i)}{\partial T_{ab}}
-
\frac{\partial C_i^{\mathrm{delay}}}{\partial T_{ab}}.
\label{eq:profit-derivative}
\end{align}

Now consider any propagation delay $T_{ab}$ for which miner $i$ is
incentive-incompatible in the mining-reward analysis, namely
$\partial \mathrm{MPR}(i)/\partial T_{ab} > 0$.
This case includes relay delays under all three tie-breaking rules, and inbound
or outbound delays whenever the corresponding coefficient in
\eqref{eq:dmpr-first-seen}--\eqref{eq:dmpr-last-generated} is positive.
Since $N^{\mathrm{blk}}>0$, $R_i>0$, $\alpha_i>0$, and
$\partial C_i^{\mathrm{delay}}/\partial T_{ab}\leq 0$, we have
\begin{align}
\frac{\partial \Pi_i}{\partial T_{ab}}
=
N^{\mathrm{blk}} R_i \alpha_i
\frac{\partial \mathrm{MPR}(i)}{\partial T_{ab}}
-
\frac{\partial C_i^{\mathrm{delay}}}{\partial T_{ab}}
>
0.
\label{eq:profit-increases-with-delay}
\end{align}
Thus, increasing $T_{ab}$ locally increases miner $i$'s real-world profit,
whereas decreasing $T_{ab}$ locally decreases it.
Consequently, whenever the mining-reward analysis shows that increasing
$T_{ab}$ is strictly profitable, namely
$\partial \mathrm{MPR}(i)/\partial T_{ab} > 0$, adding real-world profitability
factors does not restore incentive compatibility.
Most real-world factors vanish under the partial derivative with respect to the
propagation delay, and the remaining delay-dependent costs work in the same
direction as the incentive-incompatibility result.
Therefore, the incentive incompatibility identified above is not an artifact of
analyzing $\mathrm{MPR}(i)$ alone; it persists in real-world mining
profitability.

\section{Strategic Block Propagation}
\label{sec:strategic-block-propagation}

The previous section analyzed block-propagation incentives by examining the
total differentials of $\mathrm{MPR}(i)$.
This analysis gives a theoretical characterization of incentive compatibility:
it identifies when reducing a propagation delay increases or decreases a
miner's own $\mathrm{MPR}$.

This section translates those theoretical results into concrete
block-propagation strategies.
We formulate feasible strategies that miners can implement in real-world systems
such as Bitcoin and analyze the consequences of rational miner behavior.
We first consider relay-delay strategies, in which a miner does not forward
blocks generated by other miners to any other miner.
We then analyze inbound-delay and outbound-delay strategies jointly, in which a
miner temporarily behaves as if it had not yet received a block generated by
another miner or had not yet generated its own block.

\subsection{Relay-Delay Strategy}
\label{subsec:relay-delay-strategy}

We consider a \emph{non-relay strategy} for miner $i$, denoted by
$S_{\mathrm{nr}}(i)$.
Under this strategy, miner $i$ behaves in the same way as in the baseline system
until it receives a block generated by another miner.
After receiving such a block, miner $i$ updates its own local view and mines
according to the protocol, but does not relay that block to other miners.
In other words, this strategy only removes miner $i$'s contribution to the
propagation of blocks generated by other miners.

For $j \neq i$ and $k \neq i$, let $T_{jk}^{\mathrm{R}(i)}$ denote the delay
for a block generated by miner $j$ to reach miner $k$ when miner $i$ relays
blocks generated by other miners.
Let $T_{jk}^{\mathrm{NR}(i)}$ denote the corresponding delay when miner $i$
adopts $S_{\mathrm{nr}}(i)$.
We define
$\Delta_{i;jk}^{\mathrm{nr}}
:=
T_{jk}^{\mathrm{NR}(i)}-T_{jk}^{\mathrm{R}(i)}$,
and assume that $\Delta_{i;jk}^{\mathrm{nr}}$ is nonnegative for all
$j \neq i$ and $k \neq i$.
This assumption means that miner $i$'s relaying does not make blocks generated
by other miners reach other miners more slowly.
Equivalently, not relaying such blocks can only prevent propagation delays from
being improved.
If miner $i$'s relay would actually help miner $k$ receive a block generated by
miner $j$ earlier, then $\Delta_{i;jk}^{\mathrm{nr}}>0$.

Let $c_{jk}^{*}$ denote the coefficient of $dT_{jk}$ inside the braces in
\eqref{eq:dmpr-first-seen}--\eqref{eq:dmpr-last-generated} under a
tie-breaking rule $* \in \{\mathrm{FS},\mathrm{RD},\mathrm{LG}\}$.
As shown in the relay-delay analysis in the previous section,
$c_{jk}^{*}>0$ for all active miners $j$ and $k$ under all three tie-breaking
rules.
Therefore, the change in miner $i$'s mining profit rate caused by
$S_{\mathrm{nr}}(i)$ is
\begin{align}
\Delta \mathrm{MPR}_{*}(i;S_{\mathrm{nr}}(i))
&=
\frac{1}{T}
\sum_{\substack{j\in V,\,k\in V\\ j\neq i,\,k\neq i}}
c_{jk}^{*}
\Delta_{i;jk}^{\mathrm{nr}}
\geq 0 .
\label{eq:non-relay-mpr-change}
\end{align}
Moreover, if there exists at least one pair $(j,k)$ such that
$\Delta_{i;jk}^{\mathrm{nr}}>0$, then
$\Delta \mathrm{MPR}_{*}(i;S_{\mathrm{nr}}(i)) > 0$.

Thus, miner $i$ weakly benefits from not relaying blocks generated by other
miners, and strictly benefits whenever its relay would otherwise reduce at
least one relay delay.

Consequently, in a system of reward-maximizing miners, such relaying is not
incentive-compatible.
In the restricted relay game considered here, a profile in which miners do not
relay blocks generated by other miners is a Nash equilibrium.
If each miner's relay would reduce at least one relay delay, then no miner has a
profitable unilateral deviation to relaying.

\subsection{Inbound- and Outbound-Delay Strategies}
\label{subsec:inbound-outbound-delay-strategies}

We next analyze inbound-delay and outbound-delay strategies.
Unlike the relay-delay strategy, these strategies manipulate propagation delays
involving miner $i$ itself.
We formulate them as a non-cooperative game in which each miner chooses whether
to adopt delay-manipulation actions.
The utility of each miner is its mining profit rate.

For each miner $i$, we consider two delay-manipulation actions.
The first is an \emph{outbound-delay strategy}, denoted by
$S_{\mathrm{out}}(i)$.
Under this strategy, miner $i$ delays the propagation of its own newly generated
blocks by a fixed additional amount $d_i^{\mathrm{out}}>0$.
Thus, for every $k\neq i$, this strategy increases $T_{ik}$ by
$d_i^{\mathrm{out}}$.
This can be implemented by behaving toward the network, for an additional
interval $d_i^{\mathrm{out}}$, as if the block had not yet been generated.

The second is an \emph{inbound-delay strategy}, denoted by
$S_{\mathrm{in}}(i)$.
Under this strategy, after miner $i$ receives a block generated by another
miner, miner $i$ behaves for an additional interval $d_i^{\mathrm{in}}>0$ as if
that block had not yet reached it.
Thus, for every $j\neq i$, this strategy increases $T_{ji}$ by
$d_i^{\mathrm{in}}$.

We assume that the hashrate vector $(\alpha_i)_{i\in V}$ is common knowledge.
Each miner chooses only whether to adopt $S_{\mathrm{out}}(i)$ and
$S_{\mathrm{in}}(i)$.
To avoid degenerate indifference cases, we assume that the parameters are in
generic position, meaning that no miner is exactly indifferent between adopting
and not adopting any delay-manipulation action.

Let $\sigma^0$ denote the strategy profile in which no miner adopts any
inbound- or outbound-delay manipulation action.
Let $\sigma^{\mathrm{NASH}}$ denote a Nash equilibrium profile.
For a strategy profile $\sigma$, let $\mathrm{MPR}(i;\sigma)$ denote the mining
profit rate of miner $i$ under $\sigma$.
We define
\begin{align}
\mathrm{MPR}_0(i)
&:=
\mathrm{MPR}(i;\sigma^0),
\label{eq:mpr-baseline-strategy}
\\
\mathrm{MPR}_{\mathrm{NASH}}(i)
&:=
\mathrm{MPR}(i;\sigma^{\mathrm{NASH}}),
\label{eq:mpr-nash-strategy}
\\
\Delta \mathrm{MPR}_{\mathrm{NASH}}(i)
&:=
\mathrm{MPR}_{\mathrm{NASH}}(i)
-
\mathrm{MPR}_0(i).
\label{eq:nash-mpr-change}
\end{align}

Since the $\mathrm{MPR}$ expressions in
\eqref{eq:mpr-first-seen}--\eqref{eq:mpr-last-generated} are linear in the
propagation delays $T_{ij}$, the marginal effect of each delay-manipulation
strategy is independent of whether the other strategies are adopted.
Thus, each strategy can be evaluated independently.
Therefore, excluding knife-edge indifference cases, a miner adopts a strategy if
and only if the strategy's own marginal gain is positive.

\paragraph*{First-seen rule}

Under the first-seen rule, if miner $i$ adopts $S_{\mathrm{out}}(i)$, the
resulting changes in $\mathrm{MPR}$ are
\begin{align}
\Delta \mathrm{MPR}_{S_{\mathrm{out}}(i)}(i)
&=
\frac{d_i^{\mathrm{out}}}{T}
(1-\alpha_i)(2\alpha_i-1),
\label{eq:fs-sout-self-var}
\\
\Delta \mathrm{MPR}_{S_{\mathrm{out}}(i)}(j)
&=
\frac{d_i^{\mathrm{out}}}{T}
\alpha_i(1-2\alpha_i),
\qquad j\neq i.
\label{eq:fs-sout-other-var}
\end{align}
Similarly, if miner $i$ adopts $S_{\mathrm{in}}(i)$, then
\begin{align}
\Delta \mathrm{MPR}_{S_{\mathrm{in}}(i)}(i)
&=
\frac{d_i^{\mathrm{in}}}{T}
(1-\alpha_i)(2\alpha_i-1),
\label{eq:fs-sin-self-var}
\\
\Delta \mathrm{MPR}_{S_{\mathrm{in}}(i)}(j)
&=
\frac{d_i^{\mathrm{in}}}{T}
\alpha_i(1-2\alpha_i),
\qquad j\neq i.
\label{eq:fs-sin-other-var}
\end{align}
Therefore, both $S_{\mathrm{out}}(i)$ and $S_{\mathrm{in}}(i)$ are profitable
for miner $i$ if and only if $\alpha_i>1/2$.

The Nash equilibrium is characterized as follows.
If no miner has a strict majority of the total hashrate, then no miner adopts
either inbound- or outbound-delay manipulation:
$\sigma^{\mathrm{NASH}}=\sigma^0$.
Hence, $\Delta \mathrm{MPR}_{\mathrm{NASH}}(i)=0$ for all $i\in V$.

If a miner $m$ exists with $\alpha_m>1/2$, then $m$ is unique.
Miner $m$ adopts both $S_{\mathrm{out}}(m)$ and $S_{\mathrm{in}}(m)$, while all
other miners adopt neither action:
\begin{align}
\sigma^{\mathrm{NASH}}
=
\sigma^0
+
\{S_{\mathrm{out}}(m),S_{\mathrm{in}}(m)\}.
\label{eq:fs-nash-majority-profile}
\end{align}
Let $D_m^{\mathrm{FS}}:=d_m^{\mathrm{out}}+d_m^{\mathrm{in}}$.
Then the equilibrium $\mathrm{MPR}$ changes are
\begin{align}
\Delta \mathrm{MPR}_{\mathrm{NASH}}(m)
&=
\frac{D_m^{\mathrm{FS}}}{T}
(1-\alpha_m)(2\alpha_m-1),
\label{eq:fs-nash-majority-var}
\\
\Delta \mathrm{MPR}_{\mathrm{NASH}}(i)
&=
\frac{D_m^{\mathrm{FS}}}{T}
\alpha_m(1-2\alpha_m),
\qquad i\neq m.
\label{eq:fs-nash-nonmajority-var}
\end{align}

\paragraph*{Random rule}

Under the random rule, if miner $i$ adopts $S_{\mathrm{out}}(i)$, the resulting
changes in $\mathrm{MPR}$ are
\begin{align}
\Delta \mathrm{MPR}_{S_{\mathrm{out}}(i)}(i)
&=
\frac{d_i^{\mathrm{out}}}{2T}
\biggl(
4\alpha_i
-
1
-
2\alpha_i^2
-
\sum_{k\in V}\alpha_k^2
\biggr),
\label{eq:rd-sout-self-var}
\\
\Delta \mathrm{MPR}_{S_{\mathrm{out}}(i)}(j)
&=
\frac{d_i^{\mathrm{out}}}{2T}
\alpha_i(1+\alpha_j-3\alpha_i),
\qquad j\neq i.
\label{eq:rd-sout-other-var}
\end{align}
Similarly, if miner $i$ adopts $S_{\mathrm{in}}(i)$, then
\begin{align}
\Delta \mathrm{MPR}_{S_{\mathrm{in}}(i)}(i)
&=
\frac{d_i^{\mathrm{in}}}{2T}
\biggl(
4\alpha_i
-
1
-
2\alpha_i^2
-
\sum_{k\in V}\alpha_k^2
\biggr),
\label{eq:rd-sin-self-var}
\\
\Delta \mathrm{MPR}_{S_{\mathrm{in}}(i)}(j)
&=
\frac{d_i^{\mathrm{in}}}{2T}
\alpha_i(1+\alpha_j-3\alpha_i),
\qquad j\neq i.
\label{eq:rd-sin-other-var}
\end{align}

Define
\begin{align}
g_i
:=
4\alpha_i
-
1
-
2\alpha_i^2
-
\sum_{k\in V}\alpha_k^2 .
\label{eq:gi-definition}
\end{align}
Then both $S_{\mathrm{out}}(i)$ and $S_{\mathrm{in}}(i)$ are profitable for
miner $i$ if and only if $g_i>0$.

At most one miner can satisfy $g_i>0$.
Indeed, suppose two distinct miners $i$ and $j$ satisfy $g_i>0$ and $g_j>0$.
Without loss of generality, assume $\alpha_i\leq \alpha_j$.
Since $\sum_{k\in V}\alpha_k^2 \geq \alpha_i^2+\alpha_j^2$, we have
\begin{align}
g_i
&=
4\alpha_i
-
1
-
2\alpha_i^2
-
\sum_{k\in V}\alpha_k^2
\nonumber\\
&\leq
4\alpha_i
-
1
-
3\alpha_i^2
-
\alpha_j^2
\nonumber\\
&\leq
-(2\alpha_i-1)^2
\leq
0,
\label{eq:gi-at-most-one}
\end{align}
which contradicts $g_i>0$.
Therefore, there is at most one miner with profitable inbound- and
outbound-delay manipulation under the random rule.

Let $m$ be a miner with the largest hashrate share.
Since $g_i$ is strictly increasing in $\alpha_i$ for $\alpha_i<1$, the only
possible miner satisfying $g_i>0$ is the largest miner.
By the at-most-one property above, if $g_m>0$, this miner is unique.
Excluding knife-edge cases, the Nash equilibrium is
\begin{align}
\sigma^{\mathrm{NASH}}
=
\begin{cases}
\sigma^0,
&
\text{if } g_m<0,\\[1mm]
\sigma^0+
\{S_{\mathrm{out}}(m),S_{\mathrm{in}}(m)\},
&
\text{if } g_m>0.
\end{cases}
\label{eq:rd-nash-profile}
\end{align}

If $g_m<0$, no miner adopts inbound- or outbound-delay manipulation, and hence
$\Delta \mathrm{MPR}_{\mathrm{NASH}}(i)=0$ for all $i\in V$.
If $g_m>0$, let $D_m^{\mathrm{RD}}:=d_m^{\mathrm{out}}+d_m^{\mathrm{in}}$.
Then the equilibrium $\mathrm{MPR}$ changes are
\begin{align}
\Delta \mathrm{MPR}_{\mathrm{NASH}}(m)
&=
\frac{D_m^{\mathrm{RD}}}{2T}
\biggl(
4\alpha_m
-
1
-
2\alpha_m^2
-
\sum_{k\in V}\alpha_k^2
\biggr),
\label{eq:rd-nash-active-var}
\\
\Delta \mathrm{MPR}_{\mathrm{NASH}}(i)
&=
\frac{D_m^{\mathrm{RD}}}{2T}
\alpha_m(1+\alpha_i-3\alpha_m),
\qquad i\neq m.
\label{eq:rd-nash-inactive-var}
\end{align}

\paragraph*{Last-generated rule}

Under the last-generated rule, outbound-delay and inbound-delay strategies have
different profitability conditions.

If miner $i$ adopts $S_{\mathrm{out}}(i)$, the resulting changes in
$\mathrm{MPR}$ are
\begin{align}
\Delta \mathrm{MPR}_{S_{\mathrm{out}}(i)}(i)
&=
\frac{d_i^{\mathrm{out}}}{T}
(1-\alpha_i)(2\alpha_i-1),
\label{eq:lg-sout-self-var}
\\
\Delta \mathrm{MPR}_{S_{\mathrm{out}}(i)}(j)
&=
\frac{d_i^{\mathrm{out}}}{T}
\alpha_i(1-2\alpha_i),
\qquad j\neq i.
\label{eq:lg-sout-other-var}
\end{align}
Thus, $S_{\mathrm{out}}(i)$ is profitable if and only if $\alpha_i>1/2$.

If miner $i$ adopts $S_{\mathrm{in}}(i)$, the resulting changes in
$\mathrm{MPR}$ are
\begin{align}
\Delta \mathrm{MPR}_{S_{\mathrm{in}}(i)}(i)
&=
\frac{d_i^{\mathrm{in}}}{T}
\biggl(
\alpha_i
-
\sum_{k\in V}\alpha_k^2
\biggr),
\label{eq:lg-sin-self-var}
\\
\Delta \mathrm{MPR}_{S_{\mathrm{in}}(i)}(j)
&=
\frac{d_i^{\mathrm{in}}}{T}
\alpha_i(\alpha_j-\alpha_i),
\qquad j\neq i.
\label{eq:lg-sin-other-var}
\end{align}
Thus, $S_{\mathrm{in}}(i)$ is profitable if and only if
$\alpha_i>\sum_{k\in V}\alpha_k^2$.

Define the set of miners that adopt outbound-delay manipulation as
\begin{align}
X
&:=
\left\{
i\in V:
\alpha_i>\frac{1}{2}
\right\},
\label{eq:x-definition}
\end{align}
and the set of miners that adopt inbound-delay manipulation as
\begin{align}
Y
&:=
\left\{
i\in V:
\alpha_i>\sum_{k\in V}\alpha_k^2
\right\}.
\label{eq:y-definition}
\end{align}
The set $X$ contains at most one miner.
The set $Y$ is an upper set with respect to hashrate: if $i\in Y$ and
$\alpha_j>\alpha_i$, then $j\in Y$.
Moreover, under the generic-position assumption, $Y$ is nonempty.
Indeed, if $m$ is the largest miner, then
\begin{align}
\sum_{k\in V}\alpha_k^2
\leq
\alpha_m\sum_{k\in V}\alpha_k
=
\alpha_m .
\end{align}
Equality would make miner $m$ exactly indifferent between adopting and not
adopting $S_{\mathrm{in}}(m)$, which is excluded by the generic-position
assumption.
Hence $\sum_{k\in V}\alpha_k^2<\alpha_m$, and therefore $m\in Y$.
Thus, unlike the first-seen and random rules, the last-generated rule may allow
multiple miners to adopt inbound-delay manipulation.

Excluding knife-edge cases, the Nash equilibrium is
\begin{align}
\sigma^{\mathrm{NASH}}
=
\sigma^0
+
\{S_{\mathrm{out}}(i):i\in X\}
+
\{S_{\mathrm{in}}(i):i\in Y\}.
\label{eq:lg-nash-profile}
\end{align}
Equivalently, outbound-delay manipulation is adopted only by a majority miner,
if such a miner exists, whereas inbound-delay manipulation is adopted exactly
by miners whose hashrate share exceeds $\sum_{k\in V}\alpha_k^2$.

We now summarize the equilibrium $\mathrm{MPR}$ changes.
For inbound-delay manipulation, define
\begin{align}
P_Y^{\mathrm{in}}
&:=
\sum_{h\in Y}d_h^{\mathrm{in}}\alpha_h,
\label{eq:py-in-definition}
\\
Q_Y^{\mathrm{in}}
&:=
\sum_{h\in Y}d_h^{\mathrm{in}}\alpha_h^2 .
\label{eq:qy-in-definition}
\end{align}
Then the inbound-delay contribution to miner $i$'s equilibrium $\mathrm{MPR}$
change is
\begin{align}
\Delta \mathrm{MPR}_{\mathrm{NASH}}^{\mathrm{in}}(i)
=&
\frac{1}{T}
\biggl[
\alpha_i P_Y^{\mathrm{in}}
-
Q_Y^{\mathrm{in}}
\nonumber \\
&\quad
+ 
\mathbf{1}_{\{i\in Y\}}
d_i^{\mathrm{in}}
\biggl(
\alpha_i
-
\sum_{k\in V}\alpha_k^2
\biggr)
\biggr].
\label{eq:lg-nash-in-var}
\end{align}

For outbound-delay manipulation, since $X$ contains at most one miner, there
are two cases.
If $X=\emptyset$, then
$\Delta \mathrm{MPR}_{\mathrm{NASH}}^{\mathrm{out}}(i)=0$ for all $i\in V$.
If $X=\{m\}$, then
\begin{align}
\Delta \mathrm{MPR}_{\mathrm{NASH}}^{\mathrm{out}}(m)
&=
\frac{d_m^{\mathrm{out}}}{T}
(1-\alpha_m)(2\alpha_m-1),
\label{eq:lg-nash-out-majority-var}
\\
\Delta \mathrm{MPR}_{\mathrm{NASH}}^{\mathrm{out}}(i)
&=
\frac{d_m^{\mathrm{out}}}{T}
\alpha_m(1-2\alpha_m),
\qquad i\neq m.
\label{eq:lg-nash-out-nonmajority-var}
\end{align}
Therefore, for every miner $i$,
\begin{align}
\Delta \mathrm{MPR}_{\mathrm{NASH}}(i)
&=
\Delta \mathrm{MPR}_{\mathrm{NASH}}^{\mathrm{in}}(i)
+
\Delta \mathrm{MPR}_{\mathrm{NASH}}^{\mathrm{out}}(i).
\label{eq:lg-nash-total-var}
\end{align}

\section{Design Implications of Tie-Breaking Rules}
\label{sec:tie-breaking-rule-comparison}

We now compare tie-breaking rules from the perspective of protocol design.
Previous studies have often emphasized tie-breaking rules as mechanisms for
mitigating strategic mining.
For example, the last-generated rule can reduce honest miners' contribution to
an attacker's chain by exploiting the fact that a withheld block becomes older
while it is kept private.
In contrast, our results show that tie-breaking rules also play an important
role in regulating block-propagation incentives.
Specifically, among the tie-breaking rules analyzed in this paper, the
first-seen rule induces the strongest block-propagation incentives.

However, stronger propagation incentives can also imply greater competitive
pressure around propagation performance.
In a realistic network, miners with better propagation capability may obtain
higher mining profit rates, and this can worsen mining fairness.
We therefore examine the tie-breaking rules from the viewpoint of mining
fairness.
Based on this comparison, we discuss which tie-breaking rule is preferable under
different design objectives.

\subsection{The Rich-Get-Richer Effect in a Symmetric Network}
\label{subsec:symmetric-network}

We first consider a symmetric network, where $T_{ij}=0$ if $i=j$ and
$T_{ij}=d$ if $i\neq j$ for all $i,j\in V$.

Substituting this symmetric-delay setting into the analytical expressions in
\eqref{eq:mpr-first-seen}--\eqref{eq:mpr-last-generated} yields the following
common expression:
\begin{align}
\mathrm{MPR}(i)
&=
2\frac{d}{T}
\biggl(
\alpha_i
-
\sum_{j\in V}\alpha_j^2
\biggr).
\label{eq:rgr-symmetry}
\end{align}

Equation~\eqref{eq:rgr-symmetry} shows that the rich-get-richer (RGR) effect
arises even in a symmetric network: a miner with a larger hashrate share obtains
a larger mining profit rate.
Importantly, this expression is common to the first-seen, random, and
last-generated rules.
Thus, in a symmetric network, the RGR effect is caused by hashrate heterogeneity
and does not depend on the tie-breaking rule.
To compare tie-breaking rules from the viewpoint of mining fairness, we must
therefore consider asymmetric propagation delays.

\subsection{The Rich-Get-Richer Effect under Realistic Network Asymmetry}
\label{subsec:realistic-network-asymmetry}

We next compare tie-breaking rules under realistic network asymmetry.
Empirical measurements show a noticeable discrepancy between the block
propagation delay inferred from the observed fork rate and the block propagation
delay measured directly.
In fact, the propagation delay inferred from the observed fork rate is
approximately $0.09$~s, whereas the directly measured block propagation delay is
approximately $3.5$~s in Bitcoin~\cite{Calibratingtheperformance}.
Because a fork-rate-based estimate is implicitly weighted by hashrate share, it
is more strongly influenced by the propagation performance of large miners.
This suggests that, in practice, large miners enjoy a network advantage over
small miners.

Consider a small miner $i$ and a larger miner $j$, where $\alpha_i<\alpha_j$.
We evaluate whether an additional propagation disadvantage of miner $i$
increases the $\mathrm{MPR}$ gap between miner $j$ and miner $i$.
Specifically, we consider an $S_{\mathrm{out}}(i)$-type disadvantage of
magnitude $d_i^{\mathrm{out}}$ and an $S_{\mathrm{in}}(i)$-type disadvantage
of magnitude $d_i^{\mathrm{in}}$.
This does not mean that miner $i$ intentionally worsens its own propagation.
Rather, the same expressions derived for delay-manipulation strategies can be
used to evaluate the effect of realistic propagation disadvantages.

We define
\begin{align}
G_{*}^{\mathrm{out}}(i,j)
&:=
\Delta \mathrm{MPR}_{S_{\mathrm{out}}(i)}(j)
-
\Delta \mathrm{MPR}_{S_{\mathrm{out}}(i)}(i),
\label{eq:tbr-gap-def-out}
\\
G_{*}^{\mathrm{in}}(i,j)
&:=
\Delta \mathrm{MPR}_{S_{\mathrm{in}}(i)}(j)
-
\Delta \mathrm{MPR}_{S_{\mathrm{in}}(i)}(i),
\label{eq:tbr-gap-def-in}
\end{align}
where $*\in\{\mathrm{FS},\mathrm{RD},\mathrm{LG}\}$ denotes the tie-breaking
rule.
These quantities measure how much an outbound- or inbound-delay disadvantage of
miner $i$ enlarges the $\mathrm{MPR}$ gap between the larger miner $j$ and the
smaller miner $i$.

Under the first-seen rule, using \eqref{eq:fs-sout-self-var},
\eqref{eq:fs-sout-other-var}, \eqref{eq:fs-sin-self-var}, and
\eqref{eq:fs-sin-other-var}, we obtain
\begin{align}
G_{\mathrm{FS}}^{\mathrm{out}}(i,j)
&=
\frac{d_i^{\mathrm{out}}}{T}(1-2\alpha_i),
\label{eq:tbr-gap-fs-out}
\\
G_{\mathrm{FS}}^{\mathrm{in}}(i,j)
&=
\frac{d_i^{\mathrm{in}}}{T}(1-2\alpha_i).
\label{eq:tbr-gap-fs-in}
\end{align}
Thus, if $\alpha_i<1/2$, both outbound and inbound disadvantages of miner $i$
increase the $\mathrm{MPR}$ gap between miner $i$ and every larger miner $j$.

Under the random rule, using \eqref{eq:rd-sout-self-var},
\eqref{eq:rd-sout-other-var}, \eqref{eq:rd-sin-self-var}, and
\eqref{eq:rd-sin-other-var}, we obtain
\begin{align}
G_{\mathrm{RD}}^{\mathrm{out}}(i,j)
&=
\frac{d_i^{\mathrm{out}}}{2T}
\biggl(
1
-
3\alpha_i
-
\alpha_i^2
+
\alpha_i\alpha_j
+
\sum_{k\in V}\alpha_k^2
\biggr),
\label{eq:tbr-gap-rd-out}
\\
G_{\mathrm{RD}}^{\mathrm{in}}(i,j)
&=
\frac{d_i^{\mathrm{in}}}{2T}
\biggl(
1
-
3\alpha_i
-
\alpha_i^2
+
\alpha_i\alpha_j
+
\sum_{k\in V}\alpha_k^2
\biggr).
\label{eq:tbr-gap-rd-in}
\end{align}
In particular, if $\alpha_i\leq 1/3$, both quantities are positive.
Hence, the random rule also allows network disadvantages of small miners to
amplify the RGR gap.

Under the last-generated rule, using \eqref{eq:lg-sout-self-var},
\eqref{eq:lg-sout-other-var}, \eqref{eq:lg-sin-self-var}, and
\eqref{eq:lg-sin-other-var}, we obtain
\begin{align}
G_{\mathrm{LG}}^{\mathrm{out}}(i,j)
&=
\frac{d_i^{\mathrm{out}}}{T}(1-2\alpha_i),
\label{eq:tbr-gap-lg-out}
\\
G_{\mathrm{LG}}^{\mathrm{in}}(i,j)
&=
\frac{d_i^{\mathrm{in}}}{T}
\biggl(
\sum_{k\in V}\alpha_k^2
-
\alpha_i
-
\alpha_i^2
+
\alpha_i\alpha_j
\biggr).
\label{eq:tbr-gap-lg-in}
\end{align}
The outbound effect is the same as under the first-seen rule.
Moreover, if $\alpha_i<\sum_{k\in V}\alpha_k^2$ and $\alpha_j>\alpha_i$, then
$G_{\mathrm{LG}}^{\mathrm{in}}(i,j)>0$.
Thus, an $S_{\mathrm{in}}(i)$-type disadvantage of a sufficiently small miner
strictly enlarges its $\mathrm{MPR}$ gap against larger miners.

The three rules can be compared directly for each type of disadvantage.
For outbound disadvantages, \eqref{eq:tbr-gap-fs-out},
\eqref{eq:tbr-gap-rd-out}, and \eqref{eq:tbr-gap-lg-out} imply
\begin{align}
G_{\mathrm{FS}}^{\mathrm{out}}(i,j)
=
G_{\mathrm{LG}}^{\mathrm{out}}(i,j)
>
G_{\mathrm{RD}}^{\mathrm{out}}(i,j),
\end{align}
except for degenerate two-miner cases.
For inbound disadvantages, \eqref{eq:tbr-gap-fs-in},
\eqref{eq:tbr-gap-rd-in}, and \eqref{eq:tbr-gap-lg-in} imply
\begin{align}
G_{\mathrm{FS}}^{\mathrm{in}}(i,j)
>
G_{\mathrm{RD}}^{\mathrm{in}}(i,j)
>
G_{\mathrm{LG}}^{\mathrm{in}}(i,j),
\end{align}
except for degenerate two-miner cases.
Therefore, for an outbound disadvantage, the ordering of RGR-gap amplification is
$\text{first-seen}=\text{last-generated}>\text{random}$.
For an inbound disadvantage, the ordering is
$\text{first-seen}>\text{random}>\text{last-generated}$.

To compare the random and last-generated rules in aggregate, define
\begin{align}
A_i
&:=
1-2\alpha_i,
\label{eq:ai-definition}
\\
B_{ij}
&:=
\sum_{k\in V}\alpha_k^2
-
\alpha_i
-
\alpha_i^2
+
\alpha_i\alpha_j .
\label{eq:bij-definition}
\end{align}
Then,
\begin{align}
&
\bigl[
G_{\mathrm{RD}}^{\mathrm{out}}(i,j)
+
G_{\mathrm{RD}}^{\mathrm{in}}(i,j)
\bigr]
-
\bigl[
G_{\mathrm{LG}}^{\mathrm{out}}(i,j)
+
G_{\mathrm{LG}}^{\mathrm{in}}(i,j)
\bigr]
\nonumber\\
&\quad
=
\frac{d_i^{\mathrm{in}}-d_i^{\mathrm{out}}}{2T}
\bigl(A_i-B_{ij}\bigr).
\label{eq:tbr-rd-lg-aggregate-difference}
\end{align}
Since $A_i>B_{ij}$ except for degenerate two-miner cases, the aggregate ordering
between the random and last-generated rules depends on the relative sizes of
$d_i^{\mathrm{out}}$ and $d_i^{\mathrm{in}}$.
If $d_i^{\mathrm{out}}=d_i^{\mathrm{in}}$, then the random and last-generated
rules have the same aggregate effect.
If $d_i^{\mathrm{in}}>d_i^{\mathrm{out}}$, the random rule yields larger
aggregate RGR-gap amplification than the last-generated rule.
If $d_i^{\mathrm{out}}>d_i^{\mathrm{in}}$, the last-generated rule yields
larger aggregate RGR-gap amplification than the random rule.

\subsection{Design Implications}
\label{subsec:tbr-design-implications}

The comparison above reveals a fundamental trade-off.
A tie-breaking rule that gives miners stronger incentives to reduce propagation
delays also tends to make propagation advantages more valuable.
As a result, the same mechanism that encourages miners to improve propagation
can strengthen the RGR effect when propagation capabilities are uneven across
miners.

The first-seen rule provides the strongest block-propagation incentives among
the three rules: every non-majority miner has an incentive to reduce both
inbound and outbound delays.
However, in a realistic asymmetric network, it also gives the strongest
amplification of reward gaps caused by propagation disadvantages.
Thus, the first-seen rule improves propagation incentives at the cost of making
mining fairness more sensitive to propagation-performance differences.

The random rule occupies an intermediate position.
It weakens inbound and outbound propagation incentives compared with the
first-seen rule, and it also weakens RGR amplification caused by realistic
network asymmetry.
Moreover, the random rule is more favorable than the first-seen rule from the
perspective of mitigating strategic mining, because it reduces the advantage
that a miner can obtain from being received first in a fork.

The last-generated rule is attractive from a strategic-mining perspective, but
our analysis shows that it is not desirable from the viewpoint of block
propagation incentives.
It preserves the first-seen rule's outbound-delay condition, but weakens the
inbound-delay condition.
At the same time, it weakens the RGR amplification caused by inbound
disadvantages.
Thus, the last-generated rule improves fairness relative to the first-seen rule
in some asymmetric settings, but it does so by weakening miners' incentives to
receive other miners' blocks quickly.

Consequently, no tie-breaking rule is uniformly best across all design
objectives.
If the objective is to maximize block-propagation incentives, the first-seen
rule is preferable.
If the objective is to suppress RGR amplification caused by realistic network
asymmetry, the random and last-generated rules are preferable to the first-seen
rule.
If robustness against strategic mining is the primary concern, the
last-generated rule is preferable, but this comes at the cost of weaker
incentives for miners to receive other miners' blocks quickly.

\section{Limitations}
\label{sec:limitations}

Our study has several limitations.
First, we treat each mining pool as a single miner, although real mining pools
consist of geographically distributed participants coordinated by pool servers.
This abstraction ignores pool-internal delays and pool-internal stale shares and
blocks.
We expect these effects to be secondary, because pool participants typically
exchange only the minimal mining information needed for mining, which is much
smaller than full block propagation over the public Bitcoin network.
Still, explicitly modeling the distributed structure of mining pools is left for
future work.

Second, the derivation of the first-seen reward expression relies on Cond.~B.
Although this condition is natural for propagation processes, real networks may
violate it.
Removing this condition or quantifying the effect of its violations is an
important direction for future work.

Third, our simulation validation does not directly estimate $\mathrm{MPR}$ by
counting realized mining rewards over long simulation runs.
Instead, we use \textit{SimBlock} to obtain propagation-delay matrices and
hashrate distributions.
Using these network parameters, we compute $\mathrm{MPR}^{\mathrm{ref}}$ from
the underlying reward model and $\mathrm{MPR}^{\mathrm{approx}}$ from the
analytical expressions derived in this paper, and then compare the two.
Thus, the validation shows that the analytical approximation accurately
reproduces the $\mathrm{MPR}$ values computed from the underlying model under
realistic network parameters.
It should not be interpreted as a direct measurement of realized mining rewards
in a large-scale blockchain simulation.
Conducting such long-run simulations at the network scale considered in this
paper would be computationally expensive and remains future work.

Fourth, we treat propagation delays $T_{ij}$ and hashrate shares $\alpha_i$ as
fixed parameters.
A natural extension is to model them as random variables or stochastic
processes, which would enable the analysis of dynamic networks where propagation
delays and hashrate shares vary over time.

\section{Conclusion}
\label{sec:conclusion}

This paper studied the incentive compatibility of block propagation in Bitcoin.
We derived analytical reward expressions that characterize how propagation
delays, hashrate distribution, and tie-breaking rules jointly determine mining
rewards, and validated their accuracy through simulation experiments.

Using these expressions, we showed that miners have no mining-reward incentive
to reduce relay delays for blocks generated by other miners.
We also showed that incentives to reduce inbound and outbound delays become
weaker as a miner's hashrate share increases, and that a majority miner can even
benefit from delaying both sending and receiving blocks.
We further translated these incentive results into feasible block-propagation
strategies and showed that rational miners can benefit from not relaying blocks
generated by other miners and, in some cases, from strategically delaying their
own sending or receiving of blocks.

We also compared representative tie-breaking rules and revealed a trade-off:
the first-seen rule provides the strongest incentives to reduce inbound and
outbound delays, but it also makes propagation advantages more consequential for
mining rewards.
By contrast, the random and last-generated rules can mitigate such reward
unfairness at the cost of weaker propagation-improvement incentives.
Overall, our results show that Bitcoin's block propagation is not fully
incentive-compatible and that tie-breaking rules mediate the trade-off between
propagation incentives and reward distribution.

\bibliographystyle{IEEEtran}
\bibliography{hoge}

\appendix

\section{Derivation of the Analytical Reward Expressions}
\label{app:analytical-framework-derivation}

This appendix derives the analytical reward expressions used in
Section~\ref{sec:analytical-framework}.
We start from the round-based model described in Section~\ref{sec:model}.
Throughout this appendix, we use Cond.~A and omit all second- and higher-order
terms in $\varepsilon_{ij}:=T_{ij}/T$.
Let
\begin{align}
\varepsilon
:=
\max_{i,j\in V}\varepsilon_{ij}.
\end{align}
Thus, for every $i,j\in V$,
\begin{align}
F_{ij}
&=
1-\exp\left(-\frac{T_{ij}}{T}\right)
\nonumber\\
&=
\frac{T_{ij}}{T}
+
O(\varepsilon^2).
\label{eq:app-fork-probability-first-order}
\end{align}
We also recall the definitions
\begin{align}
f_{\mathrm{in}}^{(m)}(i)
&:=
\sum_{j\in V} \alpha_j^m T_{ji},
\\
f_{\mathrm{out}}^{(m)}(i)
&:=
\sum_{j\in V} \alpha_j^m T_{ij},
\\
\bar{f}
&:=
\sum_{i\in V}\sum_{j\in V}\alpha_i\alpha_j T_{ij}.
\end{align}

\subsection{Round-Start Probability}
\label{app:round-start-probability}

Let $X_r$ denote the miner that starts round $r$, and let
$q_r(i):=\Pr[X_r=i]$.
From the round transition rule, we have
\begin{align}
q_{r+1}(i)
&=
\sum_{j\in V}
\biggl\{
\alpha_i(1-F_{ji})
+
\sum_{k\in V}\alpha_k F_{jk}\alpha_i
\biggr\}
q_r(j)
\nonumber\\
&=
\alpha_i
\biggl(
1
-
\sum_{j\in V} q_r(j)F_{ji}
+
\sum_{j\in V}\sum_{k\in V}q_r(j)\alpha_k F_{jk}
\biggr).
\label{eq:app-round-start-transition}
\end{align}
Since $F_{ij}=O(\varepsilon)$, the deviation $q_r(j)-\alpha_j$ affects the
right-hand side only through terms of order $O(\varepsilon^2)$.
Therefore, to obtain the round-start probability up to first order, it is
sufficient to apply the update in \eqref{eq:app-round-start-transition} once
with $q_r(j)=\alpha_j$.
This gives
\begin{align}
q_{r+1}(i)
&=
\alpha_i
\biggl(
1
-
\sum_{j\in V}\alpha_j\frac{T_{ji}}{T}
+
\sum_{j\in V}\sum_{k\in V}
\alpha_j\alpha_k\frac{T_{jk}}{T}
\biggr)
+
O(\varepsilon^2)
\nonumber\\
&=
\alpha_i
\biggl(
1
+
\frac{\bar{f}}{T}
-
\frac{f_{\mathrm{in}}^{(1)}(i)}{T}
\biggr)
+
O(\varepsilon^2).
\end{align}
Hence, the stationary round-start probability satisfies
\begin{align}
\pi(i)
&=
\alpha_i
\biggl(
1
+
\frac{\bar{f}}{T}
-
\frac{f_{\mathrm{in}}^{(1)}(i)}{T}
\biggr)
+
O(\varepsilon^2).
\label{eq:app-pi-first-order}
\end{align}

\subsection{Reward Rate before Substituting the Tie-Breaking Rule}
\label{app:reward-rate-before-tie-breaking}

Next, we compute the reward rate of miner $i$.
If miner $i$ starts a round, miner $i$'s block receives a reward either when no
fork occurs or when a fork occurs and miner $i$'s block wins the fork.
If another miner $j$ starts a round and miner $i$ causes a fork, miner $i$'s
block receives a reward when miner $j$'s block loses the fork.
Therefore,
\begin{align}
r_i
&=
\pi(i)
\biggl(
1
-
\sum_{j\in V}\alpha_j F_{ij}
+
\sum_{j\in V}\alpha_j F_{ij}W_{ij}
\biggr)
+
\sum_{j\in V}\pi(j)\alpha_i F_{ji}(1-W_{ji})
\nonumber\\
&=
\pi(i)
-
\pi(i)\sum_{j\in V}\alpha_j F_{ij}(1-W_{ij})
+
\sum_{j\in V}\pi(j)\alpha_i F_{ji}(1-W_{ji}).
\label{eq:app-reward-rate-general}
\end{align}
In the last two terms, $F_{ij}=O(\varepsilon)$, so replacing $\pi(i)$ and
$\pi(j)$ by $\alpha_i$ and $\alpha_j$ introduces only $O(\varepsilon^2)$ error.
Substituting \eqref{eq:app-pi-first-order} into
\eqref{eq:app-reward-rate-general}, we obtain
\begin{align}
r_i
&=
\alpha_i
+
\alpha_i
\biggl(
\frac{\bar{f}}{T}
-
\frac{f_{\mathrm{in}}^{(1)}(i)}{T}
\biggr)
-
\alpha_i
\sum_{j\in V}\alpha_j F_{ij}(1-W_{ij})
\nonumber\\
&\quad
+
\alpha_i
\sum_{j\in V}\alpha_j F_{ji}(1-W_{ji})
+
O(\varepsilon^2).
\end{align}
Thus,
\begin{align}
\mathrm{MPR}(i)
&=
\frac{r_i-\alpha_i}{\alpha_i}
\nonumber\\
&=
\frac{\bar{f}-f_{\mathrm{in}}^{(1)}(i)}{T}
-
\sum_{j\in V}\alpha_j F_{ij}(1-W_{ij})
\nonumber\\
&\quad
+
\sum_{j\in V}\alpha_j F_{ji}(1-W_{ji})
+
O(\varepsilon^2)
\nonumber\\
&=
\frac{1}{T}
\biggl[
\bar{f}
-
f_{\mathrm{in}}^{(1)}(i)
-
\sum_{j\in V}\alpha_j T_{ij}(1-W_{ij})
\nonumber\\
&\quad
+
\sum_{j\in V}\alpha_j T_{ji}(1-W_{ji})
\biggr]
+
O(\varepsilon^2).
\label{eq:app-mpr-master}
\end{align}
Equation~\eqref{eq:app-mpr-master} is the common first-order expression before
substituting the tie-breaking rule.

\subsection{First-Seen Rule}
\label{app:first-seen-derivation}

We first derive the expression under the first-seen rule.
Under this rule,
\begin{align}
W_{ij}
&=
\sum_{k\in V}\alpha_k p_{i,j,k},
\end{align}
where $p_{i,j,k}$ is the probability that miner $k$ receives miner $i$'s block
before miner $j$'s block when miner $i$ starts the round and miner $j$ causes
the fork.

We use Cond.~B in this step.
Cond.~B implies
\begin{align}
T_{ik}
\leq
T_{ij}
+
T_{jk}
\end{align}
for every $i,j,k\in V$, so the case $T_{ik}>T_{ij}+T_{jk}$ in the definition of
$p_{i,j,k}$ does not occur.
If $T_{ik}<T_{jk}$, then $p_{i,j,k}=1$.
Otherwise, $T_{jk}\leq T_{ik}\leq T_{ij}+T_{jk}$, and by Cond.~A,
\begin{align}
p_{i,j,k}
&=
\frac{
\exp\left(-\frac{T_{ik}-T_{jk}}{T}\right)
-
\exp\left(-\frac{T_{ij}}{T}\right)
}{
1-\exp\left(-\frac{T_{ij}}{T}\right)
}
\nonumber\\
&=
\frac{
\left(1-\frac{T_{ik}-T_{jk}}{T}\right)
-
\left(1-\frac{T_{ij}}{T}\right)
}{
\frac{T_{ij}}{T}
}
+
O(\varepsilon)
\nonumber\\
&=
\frac{T_{ij}+T_{jk}-T_{ik}}{T_{ij}}
+
O(\varepsilon).
\end{align}
Therefore, for all $i,j,k\in V$,
\begin{align}
T_{ij}(1-p_{i,j,k})
&=
(T_{ik}-T_{jk})_+
+
O(T\varepsilon^2),
\label{eq:app-first-seen-positive-part}
\end{align}
where $(x)_+:=\max\{x,0\}$.

Using \eqref{eq:app-first-seen-positive-part}, we obtain
\begin{align}
\sum_{j\in V}\alpha_j T_{ij}(1-W_{ij})
&=
\sum_{j\in V}\sum_{k\in V}
\alpha_j\alpha_k (T_{ik}-T_{jk})_+
+
O(T\varepsilon^2),
\label{eq:app-first-seen-A}
\\
\sum_{j\in V}\alpha_j T_{ji}(1-W_{ji})
&=
\sum_{j\in V}\sum_{k\in V}
\alpha_j\alpha_k (T_{jk}-T_{ik})_+
+
O(T\varepsilon^2).
\label{eq:app-first-seen-B}
\end{align}
Subtracting \eqref{eq:app-first-seen-A} from \eqref{eq:app-first-seen-B} and
using $(x)_+-(-x)_+=x$, we get
\begin{align}
&
\sum_{j\in V}\alpha_j T_{ji}(1-W_{ji})
-
\sum_{j\in V}\alpha_j T_{ij}(1-W_{ij})
\nonumber\\
&\quad =
\sum_{j\in V}\sum_{k\in V}
\alpha_j\alpha_k
\bigl\{
(T_{jk}-T_{ik})_+
-
(T_{ik}-T_{jk})_+
\bigr\}
+
O(T\varepsilon^2)
\nonumber\\
&\quad =
\sum_{j\in V}\sum_{k\in V}
\alpha_j\alpha_k (T_{jk}-T_{ik})
+
O(T\varepsilon^2)
\nonumber\\
&\quad =
\bar{f}
-
f_{\mathrm{out}}^{(1)}(i)
+
O(T\varepsilon^2).
\end{align}
Substituting this into \eqref{eq:app-mpr-master} yields
\begin{align}
\mathrm{MPR}(i)
&=
\frac{1}{T}
\biggl(
\bar{f}
-
f_{\mathrm{in}}^{(1)}(i)
+
\bar{f}
-
f_{\mathrm{out}}^{(1)}(i)
\biggr)
+
O(\varepsilon^2)
\nonumber\\
&=
\frac{1}{T}
\biggl(
2\bar{f}
-
f_{\mathrm{in}}^{(1)}(i)
-
f_{\mathrm{out}}^{(1)}(i)
\biggr)
+
O(\varepsilon^2).
\end{align}
Omitting $O(\varepsilon^2)$ terms gives \eqref{eq:mpr-first-seen}.

\subsection{Random Rule}
\label{app:random-rule-derivation}

Under the random rule,
\begin{align}
W_{ij}
&=
\alpha_i
+
\frac{1-\alpha_i-\alpha_j}{2}.
\end{align}
Hence,
\begin{align}
1-W_{ij}
&=
\frac{1-\alpha_i+\alpha_j}{2}.
\end{align}
Therefore,
\begin{align}
\sum_{j\in V}\alpha_j T_{ij}(1-W_{ij})
&=
\frac{1-\alpha_i}{2}
f_{\mathrm{out}}^{(1)}(i)
+
\frac{1}{2}
f_{\mathrm{out}}^{(2)}(i),
\\
\sum_{j\in V}\alpha_j T_{ji}(1-W_{ji})
&=
\frac{1+\alpha_i}{2}
f_{\mathrm{in}}^{(1)}(i)
-
\frac{1}{2}
f_{\mathrm{in}}^{(2)}(i).
\end{align}
Substituting these into \eqref{eq:app-mpr-master}, we obtain
\begin{align}
\mathrm{MPR}(i)
&=
\frac{1}{T}
\biggl[
\bar{f}
-
f_{\mathrm{in}}^{(1)}(i)
-
\frac{1-\alpha_i}{2}
f_{\mathrm{out}}^{(1)}(i)
-
\frac{1}{2}
f_{\mathrm{out}}^{(2)}(i)
\nonumber\\
&\quad
+
\frac{1+\alpha_i}{2}
f_{\mathrm{in}}^{(1)}(i)
-
\frac{1}{2}
f_{\mathrm{in}}^{(2)}(i)
\biggr]
+
O(\varepsilon^2)
\nonumber\\
&=
\frac{1}{T}
\biggl[
\bar{f}
-
\frac{1-\alpha_i}{2}
\bigl(
f_{\mathrm{in}}^{(1)}(i)
\nonumber\\
&\quad
+
f_{\mathrm{out}}^{(1)}(i)
\bigr)
-
\frac{1}{2}
\bigl(
f_{\mathrm{in}}^{(2)}(i)
+
f_{\mathrm{out}}^{(2)}(i)
\bigr)
\biggr]
+
O(\varepsilon^2).
\end{align}
Omitting $O(\varepsilon^2)$ terms gives \eqref{eq:mpr-random}.

\subsection{Last-Generated Rule}
\label{app:last-generated-rule-derivation}

Under the last-generated rule,
\begin{align}
W_{ij}
&=
\alpha_i.
\end{align}
Thus,
\begin{align}
\sum_{j\in V}\alpha_j T_{ij}(1-W_{ij})
&=
(1-\alpha_i)f_{\mathrm{out}}^{(1)}(i),
\\
\sum_{j\in V}\alpha_j T_{ji}(1-W_{ji})
&=
\sum_{j\in V}\alpha_j T_{ji}(1-\alpha_j)
\nonumber\\
&=
f_{\mathrm{in}}^{(1)}(i)
-
f_{\mathrm{in}}^{(2)}(i).
\end{align}
Substituting these into \eqref{eq:app-mpr-master}, we obtain
\begin{align}
\mathrm{MPR}(i)
&=
\frac{1}{T}
\biggl[
\bar{f}
-
f_{\mathrm{in}}^{(1)}(i)
-
(1-\alpha_i)f_{\mathrm{out}}^{(1)}(i)
+
f_{\mathrm{in}}^{(1)}(i)
-
f_{\mathrm{in}}^{(2)}(i)
\biggr]
+
O(\varepsilon^2)
\nonumber\\
&=
\frac{1}{T}
\biggl[
\bar{f}
-
(1-\alpha_i)f_{\mathrm{out}}^{(1)}(i)
-
f_{\mathrm{in}}^{(2)}(i)
\biggr]
+
O(\varepsilon^2).
\end{align}
Omitting $O(\varepsilon^2)$ terms gives \eqref{eq:mpr-last-generated}.

\end{document}